# The Role of Specific Ion Effects in Ion Transport:

# The Case of Nitrate and Thiocyanate

*Kaitlin Lovering,[†] Srikanth Nayak,[†] Wei Bu,[‡] and Ahmet Uysal*[*, †]*

[†]Chemical Sciences and Engineering Division, Argonne National Laboratory, Argonne Illinois 60439, United States

[‡]NSF's ChemMatCARS, The University of Chicago, Chicago Illinois 60637, United States

AUTHOR INFORMATION

**Corresponding Author**

*E-mail: ahmet@anl.gov. Web: www.anl.gov/profile/ahmet-uysal . Phone: +1-630-252-9133

ABSTRACT

The selective transport of trivalent rare earth metals from aqueous to organic environments with the help of amphiphilic "extractants" is an industrially important process. When the amphiphilic extractant is positively charged or neutral, the coextracted background anions are not only necessary for charge balance but also have a large impact on extraction efficiency and selectivity. In particular, the opposite selectivity trends observed throughout the lanthanide series in the presence of nitrate and thiocyanate ions have not been explained. To understand the role of background anions in the phase transfer of lanthanide cations, we use a positively charged long-chain aliphatic molecule, modeling a common extractant, and gain molecular level insight into interfacial headgroup-anion interactions. By combining surface sensitive sum frequency generation spectroscopy with X-ray reflectivity and grazing incidence X-ray diffraction, we observed qualitative differences in the orientational and overall interfacial structure of nitrate and thiocyanate solutions at a positively charged Langmuir monolayer. Though nitrate adsorbs without dramatic changes to the solvation structure at the interface or the monolayer ordering, thiocyanate significantly alters the water structure and reduces monolayer ordering. We suggest that these qualitatively different adsorption trends help explain a reversal in system selectivity toward lighter or heavier lanthanides in solvent extraction systems in the presence of nitrate or thiocyanate anions.

**TOC GRAPHICS**

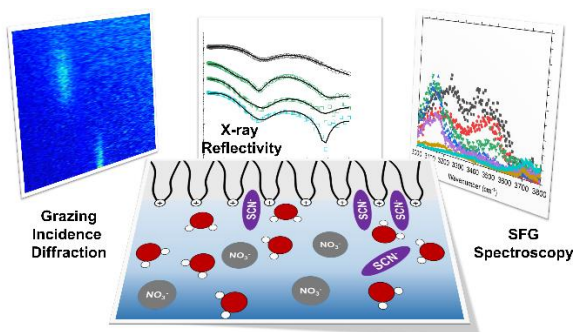

**INTRODUCTION**

The effect of ions at interfaces has been of interest since Franz Hofmeister observed the differential abilities of ions to precipitate proteins.[1-3] The implications of the Hofmeister series extend well beyond biological systems; specific ion effects (SIE) are important in a wide range of fields, including geochemistry,[4-5] atmospheric chemistry,[6-8] and chemical separations,[9-13] where ion adsorption and/or transfer at aqueous interfaces play significant roles. Due to this broad applicability, the Hofmeister series and similar empirical trends in SIE have been extensively studied to understand the effects of ions that cannot be simply explained by their charge and ionic concentration.[3-5, 14-17] The majority of these studies use the concept of "series" to explain certain physicochemical effects of ions being "more" or "less" for one ion compared to another one. Although useful in many cases, this language inevitably implies a possible single underlying mechanism to explain SIE. However, with the advancement of molecular-scale probes that can directly observe SIE at interfaces, it has become evident that such a simple and universal explanation, possibly, does not exist.[2-3, 18-19] Instead, SIE needs to be considered in a multidimensional parameter space, considering all the possible factors, such as surface functionalization and hydrophobicity, ion-ion correlations, and ion-solvent correlations that are enhanced at interfaces and in confinement.[3, 7]

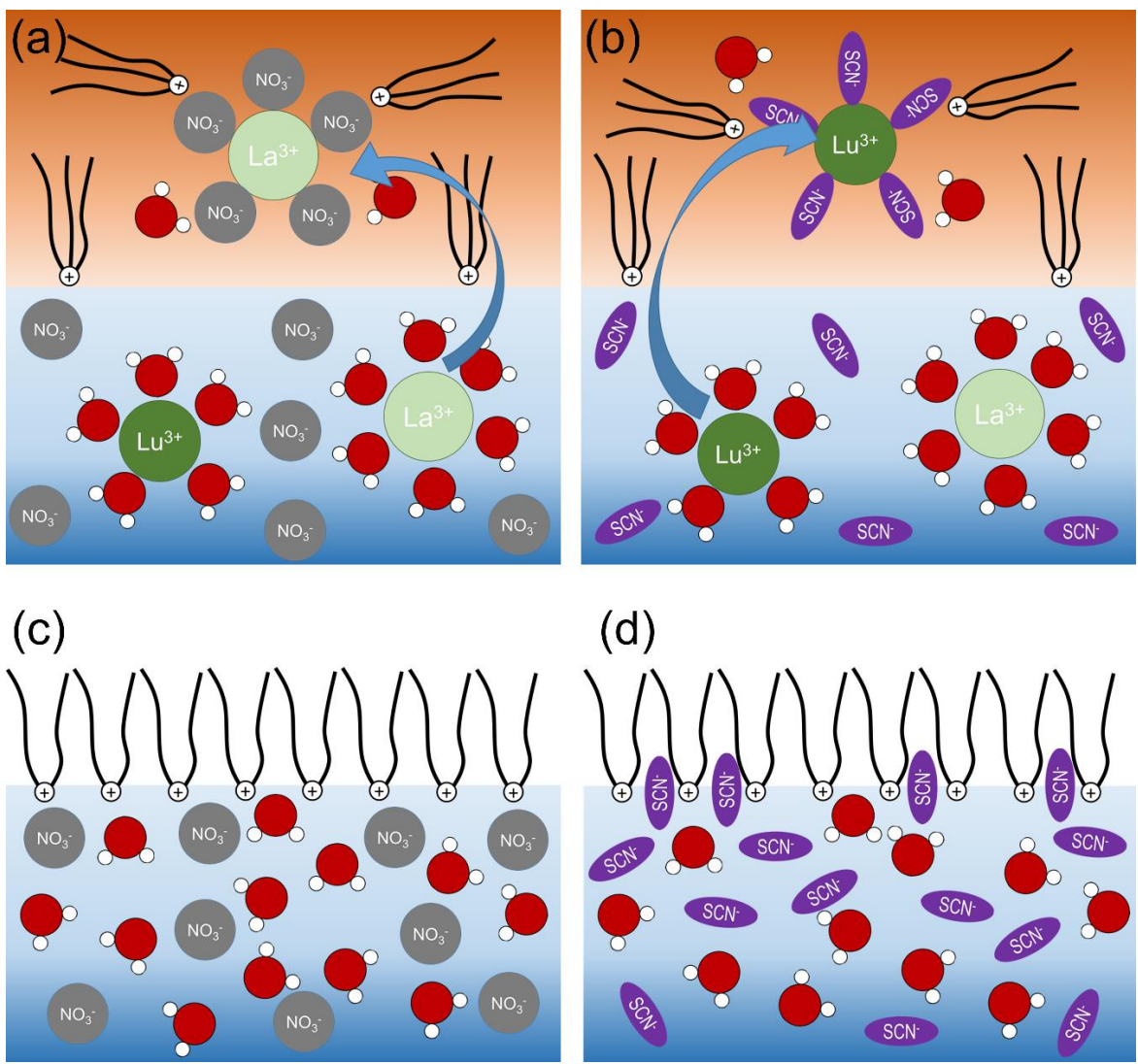

**Figure 1.** Parts a and b show schematic representations of lanthanide solvent extraction with TOMA[+]. The anion and coordination environments are based on available information in the literature. The water coordination number in the aqueous phase is known to be 9 for La[3+] and 8 for Lu[3+]. The water molecules are not intended to represent any structural information. When the background anion is $NO_3^-$, light lanthanides, represented by La, are better extracted (a). When the background anion is $SCN^-$, heavy lanthanides, represented by Lu, are better extracted (b). Parts c and d show schematic representations of model experiments designed to investigate extractant-$NO_3^-$ (c), and extractant-$SCN^-$ (d) interactions at the air/aqueous interface.

The need for a broader understanding of SIE becomes more pronounced in complex chemical processes, such as solvent extraction, a go-to chemical separations method in heavy and precious metal (including f-block elements) refinery and reprocessing.[9-12, 20-21] In solvent extraction, metal ions are selectively transferred from a mixture of ions in the aqueous phase into an organic phase with the help of amphiphilic molecules (extractants) dissolved in the organic phase (Figure 1 a, b). These extraction processes are highly affected by small changes in enthalpic and entropic contributions.[22-23] The initial and final conditions in the aqueous and organic phases can have a large impact on the distribution ratio, $D = [M^{n+}]_{org} / [M^{n+}]_{aq}$, of the target ion. While it is generally known that increasing background concentration enhances the extraction,[9-10, 22] when a positively charged or neutral extractant is used to extract a positively charged ion, the background anions can affect the selectivity in counterintuitive ways (Figure 1 a, b). For instance, when the quaternary amine methyltrioctyl ammonium (TOMA$^+$) salt is used as an extractant for lanthanides, the background anion dramatically impacts lanthanide selectivity.[24-25] Using $NO_3^-$ anions in the background makes the system selective toward the lighter lanthanides ($D_{La}/D_{Lu} \sim 300$), while systems that use $SCN^-$ anions preferentially select heavier lanthanides ($D_{Lu}/D_{La} \sim 50$). This reversal in selectivity is both interesting and industrially useful, yet the molecular-scale mechanisms responsible from this reversal have not been understood. In the only study that, to our knowledge, directly addresses this issue, extended X-ray fine structure (EXAFS) experiments probing the inner sphere coordination of all lanthanides extracted into a TOMA$^+$ $NO_3^-$ ionic liquid did not show any trend that can explain the favorable extraction of lighter lanthanides in $NO_3^-$ medium.[26]

There are multiple factors involved in the selective transfer of lanthanides from aqueous phase into the organic phase, including aqueous speciation of the ions, interfacial interactions, the diluent used as the organic phase, and the extractant-ion complexes in the organic phase.[20] The reverse micellar structures formed in the organic phase by extractants coordinating around lanthanides, anions, and water molecules are expected to determine the final thermodynamic free energy of the system and, therefore, drive the extraction process.[11, 20, 22, 27] The formation of this final structure, however, requires ion and mass transport through the aqueous/oil interface. The role of the interfacial interactions in this process are not well-understood due to the experimental inaccessibility of the interface in a real extraction. Therefore, model interfacial systems at oil/aqueous and air/aqueous interfaces have been used to understand the fundamental interactions between ions and extractants.[9-10, 28-40] For instance, recent studies showed correlations between the anion dependence in $Pr^{3+}$ extraction with a neutral extractant and the interfacial ion profiles determined by molecular dynamics (MD) simulations.[9] In another model study at the air/aqueous interface, electrostatic interactions and ion correlations were suggested to explain the higher affinity of heavier lanthanides to the negatively charged extractants.[33] However, this study was conducted at very low metal concentrations and without any background anions, which cannot be easily translated to industrial process conditions.

The extraction of rare earths by $TOMA^+$ clearly shows that the effects of background anions are highly important in understanding the extraction process. Here, we focus on a very simple and well-controlled model system at the air/aqueous interface (Figure 1c, d) to investigate the interfacial interactions between positively charged extractants and the anions present in the solution as a first step to understand this complex process.

Interface specific experimental probes, such as vibrational sum frequency generation (VSFG) spectroscopy,[38-40] X-ray reflectivity (XR),[28-32] grazing incidence X-ray diffraction (GID),[34-35, 41] second harmonic generation (SHG) spectroscopy,[36] and neutron reflectivity (NR)[37] have been shown to be very useful to investigate interfacial amphiphile ion interactions. The combination of VSFG, XR, and GID is especially useful in elucidating complementary aspects of the interfacial solute and solvent structures.[32, 38] VSFG provides vibrational structure of molecules at interfaces because, as a second-order nonlinear response, under the dipole approximation, VSFG only occurs where there is a break in inversion symmetry.[42-45] XR measures the electron density profile (EDP) at the air/aqueous interface with subnanometer resolution and provides direct structural information.[46-47] GID probes the in-plane ordering of extractants and ions and elucidates the interactions between ions and the extractants.[47-48]

In this paper, we present an experimental study of qualitative differences between $NO_3^-$ and $SCN^-$ interactions with a model quaternary amine extractant at the air/aqueous interface (Figure 1 c, d). The combination of VSFG, XR, and GID studies clearly shows that the differences between $SCN^-$ and $NO_3^-$ surface activity go beyond a simple Hofmeister series explanation. Interestingly, $NO_3^-$ and $SCN^-$ are both described as chaotropes in the literature, having relatively weak hydration free energies of -300 and -280 kJ/mol respectively.[49] A simple evaluation based on the Hofmeister series would not suggest any significant difference between them. Indeed, the work by Sung *et al.* effectively establishes that the surface activity of halide anions follows the Hofmeister series at the DPTAP interface.[15] Our experiments indicate that while $NO_3^-$ adsorption as a function of bulk $NO_3^-$ concentration creates predictable effects on interfacial water structure, similar to the halide anions, $SCN^-$ ions, which appear to adsorb in two different populations, cause interfacial water to reorganize and significantly affect the extractant structure. We suggest that these qualitatively different adsorption trends may help explain the preference of lighter or heavier lanthanides in solvent extraction systems in the presence of $NO_3^-$ or $SCN^-$ anions, respectively.

## EXPERIMENTAL METHODS

The experimental techniques have been discussed in detail previously[32, 38, 47] and the key points are described here for completeness.

**Sample Preparation.** Sodium thiocyanate (NaSCN, $\geq 98\%$), sodium nitrate (NaNO$_3$, $\geq 99\%$), and HPLC-grade chloroform (CHCl$_3$, $\geq 99.9\%$) were purchased from Sigma-Aldrich. 1,2-Dipalmitoyl-3-trimethylammonium-propane (DPTAP$^+$) chloride salt was purchased in the powder form from Avanti Polar Lipids and stored at $-20°$ C. All chemicals were used as received. Solutions of 0.25 mM DPTAP$^+$-Cl$^-$ in CHCl$_3$ were prepared and stored at 4° C. DPTAP$^+$ solutions were discarded after 10 days and were not subject to any freeze–thaw cycles. All subphase solutions were 30 mL. Ultrapure water with a resistivity of 18.2 M$\Omega$ cm (Millipore, Synergy Water Purification System) was used to prepare each subphase.

Langmuir monolayer samples were prepared in a $60 \times 20$ mm$^2$ flat-form polytetrafluoroethylene dish. The Langmuir monolayer was prepared using the dropwise addition of 0.25 mM DPTAP$^+$ in CHCl$_3$ from a 1 μL Hamilton syringe. A Nima pressure sensor using a chromatography paper Wilhelmy plate was used to measure the surface pressure. Experiments were performed at room temperature ($\sim 20$ °C) with a surface pressure between 12 and 8 mN/m. Similar systems studied by SFG and X-ray techniques indicate that the DPTAP$^+$ monolayer is stable under these conditions for the duration of the experiments.[32, 38]

**VSFG Experiments.** VSFG is a second-order nonlinear technique that probes molecules in noncentrosymmetric environments, such as interfaces. The details of VSFG are discussed elsewhere, however, the VSFG signal intensity can provide information on vibrational environment, molecular orientation, and number density of oscillators.[50-51] The vibrational response of the system is the second order susceptibility, $\chi^{(2)}$. $\chi^{(2)}$ is an average of the molecular hyperpolarizabilities of the component oscillators and is a third rank tensor containing 27 elements that can be written, in laboratory coordinates, in the form $\chi_{ijk}^{(2)}$.[52-54] At the majority of interfaces, which are isotropic, only seven of the elements are active. Furthermore, the polarization configurations do not concomitantly probe each of these seven active elements, and a comparison of the VSFG intensities under different polarization combinations gives insight into the orientation of the molecules at the interface.[52-53]

Experiments were carried out using an EKSPLA SFG spectrometer. Details of the system have been published previously. In brief, a 1064 nm beam is split and frequency is doubled to 532 nm. One of the 532 nm beams is directly used to probe the sample while the remaining beams are parametrically combined to generate a tunable IR beam. The beams are overlapped in space and time in a reflection geometry; the visible and IR excitation angles, with respect to the sample normal, are $\theta_{vis} = 60°$ and $\theta_{IR} = 55°$, respectively. The 532 nm polarization is adjusted with a $\lambda/2$ waveplate, and the SFG signal is selected using a Glan polarizer. After the sample, the SFG signal is directed to a monochromator (Sol, MS2001) and collected with a photomultiplier tube (Hamamatsu, R7899).

A motorized piezoelectric rotation stage (ThorLabs, ELL8K) rotates the sample after three frequency steps to avoid sample damage. Each spectrum was collected with a 4 cm$^{-1}$ increment over the range 2800–3800 cm$^{-1}$ and averaged 50 laser shots per point. The spectra were normalized against the SFG spectrum of z-cut quartz. VSFG data were collected with both the SSP and PPP polarization conditions, where S(P)-SFG signal; S(P)-VIS; and P(S)-IR. The electric field of P-polarized light is parallel to the plane of incidence, and the electric field of S-polarized light is perpendicular to the plane of incidence.

**Synchrotron X-ray Experiments.** Experiments were done at Sector 15-ID-C, NSF's ChemMatCARS, of the Advanced Photon Source at Argonne National Laboratory. The X-ray energy was tuned to 17 keV. Two pairs of motorized slits set the incident beam size to 2 mm horizontally and 0.02 mm vertically. A Pilatus 100 K area detector records the scattered X-ray signal. The sample chamber was purged with helium to reduce the beam damage and the background scattering. The sample was shifted perpendicular to the beam, periodically, to avoid any beam damage due to long X-ray exposure.

**XR Measurements**. The specularly reflected X-ray intensity was recorded as a function of the vertical momentum transfer $|\vec{q}| = (4\pi/\lambda)\sin(2\theta/2)$, where $\lambda$ (0.73 Å at 17 keV) is the wavelength and $\theta$ is the incidence angle. The electron densities of the films were modeled by two slabs, one for the tail region and the other for the headgroup plus the adsorbed ions, or by three slabs when extensive monolayer modifications occurred due to ion interactions (see discussion). The thickness, electron density, and roughness of these layers are determined by least-squares fitting of the XR data to the calculated XR curves according to the Parratt formalism (Table S1).

**GID Measurements.** The X-ray energy was fixed at 17 keV. The incidence X-ray angle was fixed to 0.019 Å$^{-1}$, and the detector was moved in the plane of the water surface to record the diffraction patterns. Only a 3 pixel wide (~0.5 mm) stripe of the area detector was used in a diffraction pattern reconstruction to obtain high $q_{xy}$ resolution; $q_z$ resolution was defined by the pixel size (172 μm). The peak positions were determined from linear plots obtained by the vertical and horizontal integration of the diffraction patterns. The molecular areas and the tilt angles are calculated from these peak positions.

**RESULTS AND DISCUSSION**

We use 1,2-dipalmitoyl-3-trimethylammonium-propane chloride (DPTAP[+] Cl[-]) monolayers to model TOMA[+] (Figure 1c, d). DPTAP[+] has a longer chain and hence forms a stable Langmuir monolayer at the air/aqueous interface. DPTAP[+] has previously been used to study anion adsorption at air/aqueous interface[14-16, 55] as well as to mimic TOMA[+] interactions with platinum group metals at the air/aqueous interface.[32, 38] Details of the measurements are explained in the Experimental section and the Supporting Information.

VSFG spectra probing the interfacial water in Figure 2 show significant differences between the interfacial region populated by NO$_3^-$ anions and the region populated by SCN[-] anions. The VSFG spectra of four concentrations of NaNO$_3$ show a uniform decrease of intensity (Figure 2a). The reduction in water signal can be qualitatively explained by the Guoy-Chapman theory, which describes how charged species interact with an oppositely charged surface.[56-57] As the bulk ion concentration increases, more ions approach the surface and shield the surface charge, reducing the ability of the surface to induce order in the aqueous subphase.[58-59] The VSFG spectra, especially at low ionic strength and/or high surface charge, can include third-order contributions due to the static field.[57-58] Interference between second- and third-order susceptibility tensors can attenuate peak position and intensity.[60-62] In the NO$_3^-$ system, increased charge shielding does not result in peak shifts, only decreased overall VSFG intensity, indicative of a reduction of the number of water molecules contributing to the VSFG signal.[63-64]

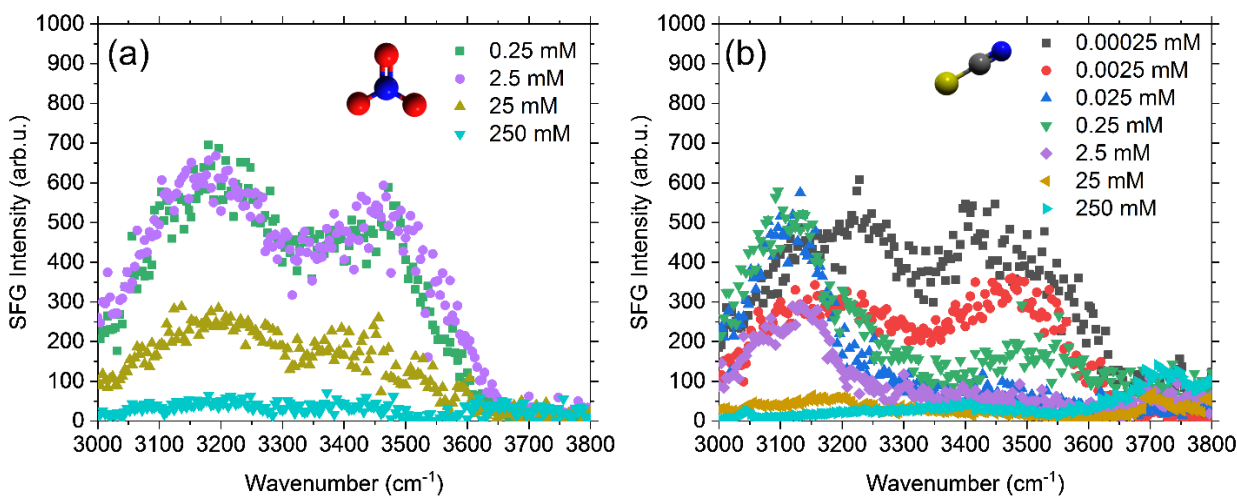

**Figure 2.** OH stretching region of (a) NaNO₃ solutions and (b) NaSCN solutions in contact with DPTAP⁺ monolayers in a SSP polarization combination. In part a, the SFG intensity decreases uniformly with concentration, while in part b the SCN ions apparently lead to more substantive changes at the interface.

The SCN⁻-containing interface shows three major differences (Figure 2b) in comparison to the NO₃⁻ samples. First, in order to observe the bimodal shape of a typical water region, it is necessary to use concentrations in the micromolar range, approximately $10^3$ times less than what was used in the NO₃⁻ system. Second, at 0.025 mM, water peak loses its bimodal character; the 3500 cm⁻¹ peak almost disappears, while the 3200 cm⁻¹ peak redshifts to 3100 cm⁻¹ and becomes narrower. Finally, as the signal intensity from the bonded OH region decreases, a free OH peak appears, indicating disruption in the monolayer surface coverage. The shift and narrowing of the hydrogen-bonded water peak in Figure 2b is strikingly different from that in Figure 2 a and the ionic impacts observed at other surfaces.[4-6, 14-16] It is important to note that the comparisons made at the 0.25 mM to 0.25 M concentration range for both anions suggest that the significant differences between the SCN⁻ and NO₃⁻ samples are mainly due to the $\chi^{(2)}$ effects, since $\chi^{(3)}$ effects are expected to be similar at the same concentration.

VSFG is sensitive to not only the number of oscillators but also their orientation. The DPTAP$^+$/pure water interface provides an example of the importance of the molecular orientation of the oscillators; there is a dip at 2950 cm$^{-1}$ due to destructive interference between OH and CH oscillators pointing in opposite directions.[15, 65] SCN$^-$ ions are known to lead to charge reversal at similar interfaces and at intermediate concentrations, it is possible that the apparent red-shift is, at least partially, due to constructive interference between the CH and OH oscillators.[18, 55] As SCN$^-$ concentration increases all OH stretches disappear.[18, 55] At 250 mM, a stretch from the quaternary amine headgroup, usually hidden by OH stretches, becomes visible (Figure 2b, cyan triangles, 3050 cm$^{-1}$).[66]

The Hofmeister series predicts SCN$^-$ anions to have a larger impact at the interface than NO$_3^-$ anions, and this trend is well established in VSFG literature.[5, 14] These SIE are typically treated as differences in magnitude of the adsorption energy. However, Figure 2 suggests that in addition to increased surface activity leading to more charge shielding at lower concentrations, the adsorption pathway of SCN$^-$ in the free energy landscape is different than that of NO$_3^-$.[18] The distinctive behavior of SCN$^-$ was also suggested in work by Leontidis and co-workers, where they correlated the effects of anions on the surface pressure of a zwitterionic surfactant with their thermochemical radii and showed SCN$^-$ is an outlier, not following the expected trend.[3, 67]

In order to better understand the adsorption behavior leading to a qualitatively different water structure, the CN stretch of SCN$^-$ was probed using SSP and PPP polarization configurations (Figure 3). There was some spot dependence in the signal intensity (consistent with the appearance of the free OH peak in Figure 2b), and the spectra shown in Figure 3 are an average of multiple scans. The standard deviations around the average are shown in Figure S2. The SSP signal does not vary from the smallest concentration up to the highest concentration. The PPP signal increases with the increasing bulk SCN⁻ concentration, appears to saturate around 2.5 mM, and does not significantly change up to 250 mM. The SSP configuration probes the $yyz$ component of the $\chi^{(2)}$ tensor while PPP probes the combination of the $zzz$, $yyz$, $yzy$, and $zyy$ components.[53] In the SSP configuration, the signal does not significantly change, despite increasing the bulk concentration by a factor of $10^6$. This indicates that the SCN⁻ population that contributes to this signal saturates the surface at the lowest concentration probed. The CN stretch probed in the PPP configuration increases with concentration. Particularly at higher concentrations, the SSP:PPP intensity ratio is very small (~1/8), indicating that the ions are orienting themselves more parallel to the surface normal compared to the angle observed at lower concentrations.[68] Combining this information, we surmise that either SCN⁻ ions saturate at one orientation and all subsequent ions must adsorb perpendicularly to the interface or, as more ions are added to the interface, they are forced to reorient. Regardless, at higher concentrations, the majority of SCN⁻ anions are adsorbed with a relatively perpendicular orientation with respect to the interface.

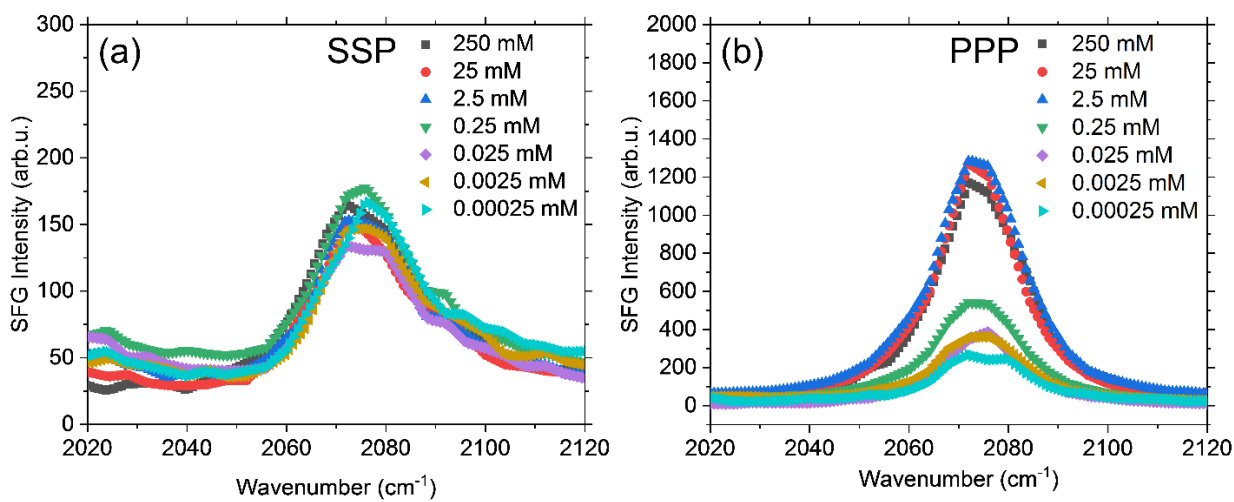

**Figure 3.** CN stretch of the SCN ions shown in (a) SSP and (b) PPP polarization combinations. Notably, the peak of the PPP is significantly more intense than the SSP configuration. Additionally, the SSP intensity is relatively constant while the PPP signal shows an increase with increasing ion concentration. There is some spot dependence in the SFG intensity of this peak, and the curves are an average of multiple scans. The standard deviation of the averages are shown in Supporting Information (Figure S2).

An investigation of the interfacial EDPs and the response of the DPTAP$^+$ monolayer to the adsorbed ions provide more insights about the qualitative differences between NO$_3^-$ and SCN$^-$ adsorptions. Figure 4 a (symbols) shows the XR data collected at two different concentrations with NO$_3^-$ and SCN$^-$ ions in the subphase. The data were fit according to Parratt formalism, modeling the interface as uniform electron density layers ("boxes"), with error function interfaces between them.[32, 46-47, 69] The model parameters include the thickness and the electron density of the boxes as well as the roughness between them (Table S1). The known parameters, such as the bulk aqueous density, are fixed and other parameters are allowed to float within reasonable limits. The details of the fits are described in the Methods section and in the Supporting Information. The lower concentrations were chosen so that the bimodal water signal is still visible in VSFG measurements, corresponding to 0.25 µM and 0.25 mM for SCN$^-$ and NO$_3^-$, respectively (Figure 2). Though the VSFG spectra of the OH stretching region of these two subphases suggests that the anions interact similarly with the monolayer, the EDPs are significantly different. The EDP for DPTAP$^+$ on 0.25 µM SCN$^-$ is identical to the one obtained at pure water surface,[15] suggesting minimal interaction between SCN$^-$ and DPTAP$^+$. However, 0.25 mM NO$_3^-$ data show significant distortion of the monolayer and significant ion adsorption at the interface, a very rough and low density tail group region is needed to fit the data, which is comparable to DPTAP$^+$/Cl$^-$ and I$^-$ interfaces reported previously.[15] It is important to note that VSFG measurement of CN stretch show the presence of SCN$^-$ ions even at the lowest concentration (Figure 3), yet XR show that they are not interacting with DPTAP$^+$. VSFG is sensitive to several tens of nanometers at the interface, as long as the inversion symmetry is broken. XR is only sensitive to the gradient of the EDP and therefore cannot detect small amounts of $SCN^-$ if they are not forming a well-defined layer at the interface. Combination of this information suggests that at 0.25 µM concentration, $SCN^-$ ions are not directly interacting with $DPTAP^+$, yet they create a population with certain orientational order.

Increasing the bulk $SCN^-$ concentration to 0.25 mM significantly changes the $DPTAP^+$ structure, and fitting requires a three-box model (two for the headgroup/ion region and one for the tail region). While the increasing length and roughness of the tail group region looks comparable to $NO_3^-$ data, the regions corresponding to the headgroup/ion layers appear to be sharper for 0.25 mM $SCN^-$ (Figure 4b). In contrast, increasing the bulk $NO_3^-$ 3 orders of magnitude to 0.25 M does not change the monolayer structure significantly, instead mostly increasing the density at the headgroup/ion layer (Figure 4b). These results are in good agreement with VSFG results that show no qualitative difference between low and high concentration $NO_3^-$ systems but a qualitatively different water structure for low and high concentration $SCN^-$ systems (Figure 2).

The GID results, elucidating the in-plane organization of $DPTAP^+$ molecules, corroborate the XR observations. For all concentrations of $SCN^-$ and $NO_3^-$ solutions, we observe one in-plane peak and one out-of-plane peak, corresponding to a distorted hexagonal packing of alkyl tails with a tilt toward the nearest neighbor (NN) (Figure 4 b, c).[47-48] For 0.25 µM $SCN^-$, the tilt angle is 37° from the surface normal and the molecular area per $DPTAP^+$ molecule is 51 $Å^2$. As the bulk $SCN^-$ concentration is increased to 0.25 mM, the in-plane peak shifts to 1.71 $Å^{-1}$ as a result of significant change in the monolayer structure. The tilt angle becomes 25° from the surface normal, and the molecular area decreases to 43 $Å^2$ per $DPTAP^+$. This change of tail group orientation is also observed with VSFG of the CH region in the PPP configuration (Figure S3). Though the water response is different, the NO$_3^-$ solutions of 0.25 mM and 0.25 M show diffraction peaks and tilt angles equivalent to those of the higher concentration solution of SCN$^-$ (Figure 4 d).

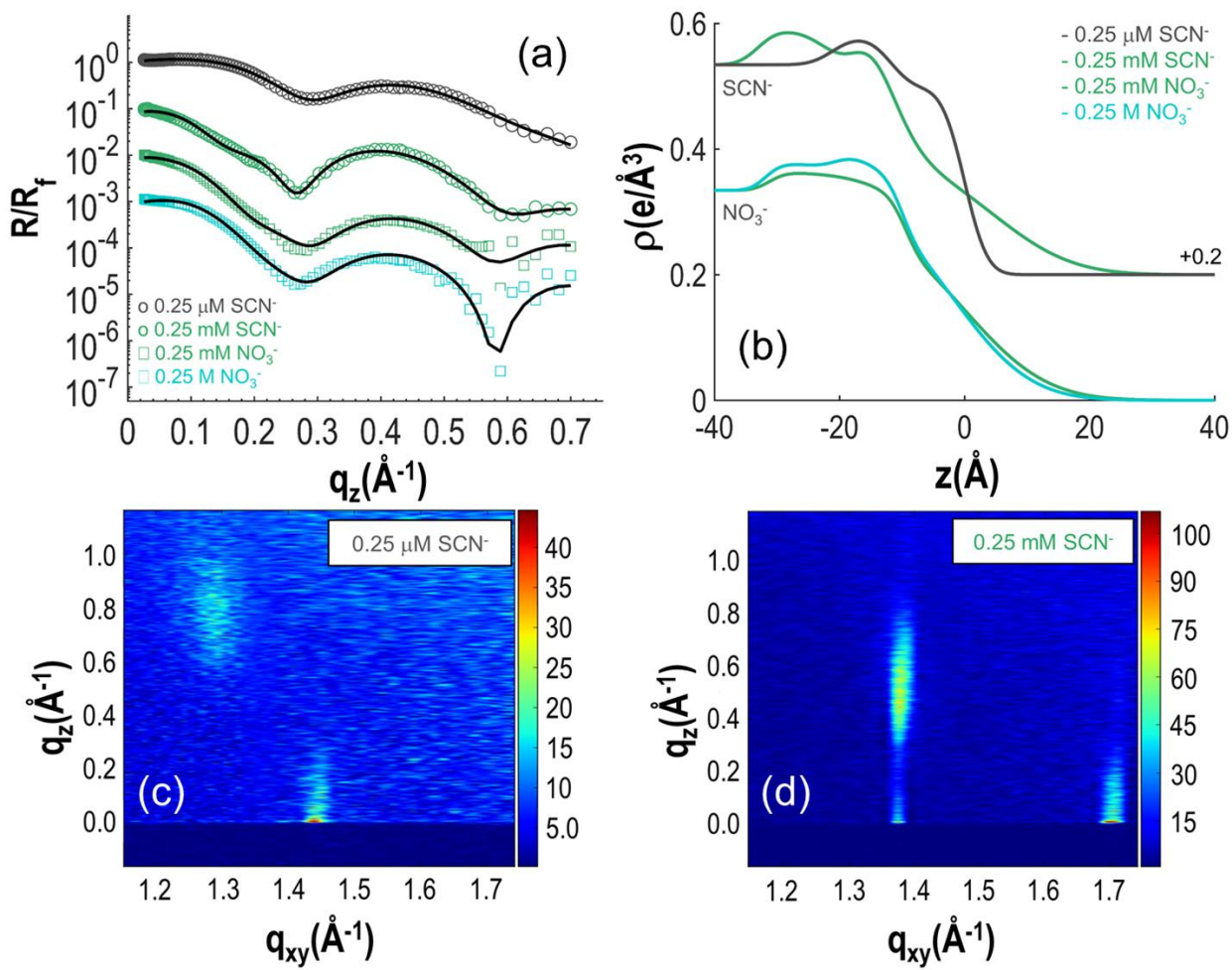

**Figure 4.** (a) XR data and fits for DPTAP$^+$/aqueous interface with 0.25 μM SCN$^-$ and 0.25 mM SCN$^-$, 0.25 mM NO$_3^-$, and 0.25 M NO$_3^-$ concentrations in the subphase. A two-box model was sufficient to fit 0.25 μM SCN$^-$ and 0.25 mM NO$_3^-$ data, while three-box models were used for the other two samples. Each data set is scaled by a decade for clarity. (b) EDPs derived from the fits shown in part a. The inflection point in the tail group region is selected as zero. SCN$^-$ EDPs are shifted +0.2 for clarity. (c-d) GID data from 0.25 μM and 0.25 mM SCN$^-$ samples, respectively. GID data for both NO$_3^-$ samples look identical to that in part d (not shown).

These results clearly show the quantitative and qualitative differences in $SCN^-$ and $NO_3^-$ adsorption at a positively charged, soft interface. One interesting result arising from the comparison of XR and VSFG data is that $NO_3^-$ ions significantly affect the $DPTAP^+$ structure without significantly altering the overall interfacial water structure (Figures 2a and 4b, 0.25 mM $NO_3^-$ data). In contrast, when $SCN^-$ ions have a similar effect on $DPTAP^+$, the interfacial water structure is altered significantly; a single red-shifted VSFG peak is observed instead of a typical bimodal shape (Figures 2b and 4b, 0.25 mM $SCN^-$ data). Also, at higher concentrations, $SCN^-$ ions appear to be more incorporated into $DPTAP^+$ monolayer,[67] causing a significant distortion in the monolayer coverage and the appearance of a free OH peak at high $SCN^-$ concentrations (Figure 2b) and decreasing the in-plane uniformity of the monolayer.

During the solvent extraction, $M^{3+}$ ions are expected to partially lose their hydration shell because they are moving into a hydrophobic phase.[11, 20, 22, 27] However, it is known that the amount of water transferred with metals is not negligible, though they may not be directly coordinating the metal ions.[20] A recent study showed a strong correlation between the amount of water in the organic phase and the distribution ratio of rare earth ions, without any evidence of water in the first coordination shell.[27] Our results show that $NO_3^-$ ions can interact with $DPTAP^+$ without significantly affecting the interfacial hydrogen bonding network. This may lead to a more efficient transport of water into the organic phase and therefore transport of $M^{3+}$ ions without a significant loss of their hydration shells. In contrast, $SCN^-$ ions appear to replace interfacial water significantly and may force extraction to happen through a different mechanism, possibly dominated by electrostatic interactions. It is reasonable to expect heavier lanthanides to be better extracted when the electrostatic interactions become more important.[33]

These results also suggest that the reverse micellar structures formed in the organic phase in the presence of $NO_3^-$ or $SCN^-$ ions are significantly different than each other. Although there has not been any systematic and comparative investigation of these structures, in recent solvent extraction experiments of $Eu^{3+}$ with Aliquat-336, a commercial extractant very similar to $TOMA^+$, Knight *et al.* showed that the aggregation numbers (extractant per reverse micelle in the organic phase) are significantly different in $NO_3^-$ (5.4) and $SCN^-$ (8.5) media.[22] The charge balance of reverse micelles, then, strongly suggests that the number of $SCN^-$ ions that stabilize the $M^{3+}$ ions extracted from the $SCN^-$ medium is higher than the number of $NO_3^-$ ions that stabilize the $M^{3+}$ ions extracted from the $NO_3^-$ medium. However, the elucidation of the details of these structures requires small-angle X-ray scattering investigations.

The combination of X-ray and laser-based surface sensitive measurements allows the disambiguation of the impact of ions on monolayer and water structure. GID measurements show that $SCN^-$ and $NO_3^-$ have a similar impact on the in-plane ordering of $DPTAP^+$, although the appearance of the free OH indicates that $SCN^-$ decreases the crystalline island coverage. VSFG spectra also clearly show a different response in the hydrogen-bonded OH region and, in combination with XR spectra, indicate that when $SCN^-$ directly binds to the $DPTAP^+$ surface, the water structure is significantly altered. At moderate concentrations, $SCN^-$ ions lead to a reorientation of the water molecules at the surface; at higher $SCN^-$ concentrations, the interface is largely dehydrated. As the selective transport of lanthanides across the aqueous/organic interface cannot be explained by electrostatic attraction between the $DPTAP^+$ headgroup and cation, these ion-ion and ion-solvent interactions at the interface are expected to play an important role in the transport of trivalent rare earth ions from an aqueous phase into an organic phase in the presence of highly concentrated $SCN^-$ or $NO_3^-$ ions (Figure 1a, b).

**CONCLUSIONS**

In summary, we have identified an interesting problem in amphiphile and anion assisted interfacial ion transport of lanthanides, as the character of the anion ($NO_3^-$ or $SCN^-$) results in preferred extraction of (lighter or heavier, respectively) lanthanides. A thorough understanding of this process requires a detailed investigation of aqueous phase speciation, organic phase reverse micelles, and the interfacial interactions, which cannot be managed in a single publication. In this study, we investigated the very fundamental interfacial behavior of $NO_3^-$ and $SCN^-$ ions, without any metal ions, and showed that it is possible to identify qualitative differences that may lead to dramatic differences in the real process. Further studies elucidating other aspects of this multifaceted process will provide important molecular scale details about ion transport in industrially relevant complex fluids.

ASSOCIATED CONTENT

**Supporting Information.**

Supporting Information is available free of charge at

Additional VSFG data and XR fitting parameters (PDF)

AUTHOR INFORMATION

**Corresponding Author**

*E-mail: ahmet@anl.gov. Web: www.ahmet-uysal.com. Phone: +1-630-252-9133

**Notes**

The authors declare no competing financial interests.

ACKNOWLEDGMENTS

This work is supported by the U.S. Department of Energy, Office of Basic Energy Science, Division of Chemical Sciences, Geosciences, and Biosciences, under contract DE-AC02-06CH11357. NSF's ChemMatCARS Sector 15 is principally supported by the Divisions of Chemistry (CHE) and Materials Research (DMR), National Science Foundation, under grant number NSF/CHE-1834750.  Use of the Advanced Photon Source, an Office of Science User Facility operated for the U.S. Department of Energy (DOE) Office of Science by Argonne National Laboratory, was supported by the U.S. DOE under Contract No. DE-AC02-06CH11357.